\documentclass[%
 reprint,
 prl,
]{revtex4-2}

\usepackage{graphicx}
\usepackage{dcolumn}
\usepackage{bm}
\usepackage{amsmath}

\begin{document}


\title{Autocorrelation Invariance Property of Chaos for Wireless Communication}


\author{Hui-Ping Yin}
\affiliation{Shaanxi Key Laboratory of Complex System Control and Intelligent Information Processing, Xi'an University of Technology, Xi'an 710048, China}
\author{Hai-Peng Ren}
\affiliation{Shaanxi Key Laboratory of Complex System Control and Intelligent Information Processing, Xi'an University of Technology, Xi'an 710048, China}
\author{Celso Grebogi}
\affiliation{Institute for Complex Systems and Mathematical Biology, University of Aberdeen AB24 3UE, United Kingdom}



\begin{abstract}
A new feature of the chaotic signal generated by chaotic shape-forming filter (CSF) is uncovered in this work. We find that, the autocorrelation function (ACF) of the transmitting signal generated by CSF keeps the same as that of the base function of CSF, no matter what information is encoded. We derive the analytical equation to describe the relation between the ACF of the received signal and the wireless channel parameters using the ACF of the transmitted signal as prior knowledge revealed by the finding in this work. This new property can be utilized together with different wireless communication systems to improve the system performance. Specially, to demonstrate the improvement, channel state information (CSI) is identified using the chaotic baseband wireless communication as a paradigm. Two significant benefits by using the new property are 1) the CSI can be identified without the probe information known to the receiver as done in the conventional wireless communication systems, which improves the bandwidth efficiency, especially in the time-varying channel; 2) the correlation operation is insensitive to the channel noise, which improves the identification accuracy as compared to the commonly used methods.
\end{abstract}


\maketitle


Wireless communication is widely used in daily life from Wi-Fi \cite{crow1997ieee} and mobile phone \cite{katz2002perpetual} to internet of things \cite{ashton2009internet}, satellite communication \cite{elbert2008introduction}, and underwater communication \cite{bai2019digital}. Although various kinds of advanced technology have been developed for wireless communication systems to support the fastest growing fields such as mobile phone from G1 to G5, G6 and internet of things applications, the even higher communication performance demand calls for more advances in basic theory and technology.

As a natural signal with noise-like spectrum and sensitive dependence on initial conditions, the chaotic signal has been thought to be suitable for application in communication, since it was recognized as a controllable and synchronizable signal in the seminal work in Refs. \cite{ott1990controlling,pecora1990synchronization,konnur1996equivalence}. At the very beginning, chaos applications in communication were mainly focused on secure communication \cite{pecora1990synchronization} and spread spectrum \cite{halle1993spread,parlitz1994robust}. Among them, secure communication with chaos was extensively investigated and proved to lack the necessary security \cite{cuomo1993synchronization,feki2003adaptive,ren2015secure,perez1995extracting,yang1998breaking,yang2004survey,hai2008breaking}, then this branch of chaos application gradually diminished. Another branch of chaos application in communication evolved into the direct spread spectrum sequence (DSSS) by replacing the conventional quasi-random sequence with chaotic signal and the differential chaotic shift keying (DCSK) by using chaotic signal as the reference signal. The latter one has developed into a local wireless network standard IEEE 802.15.6 \cite{Ieee2012802}.

Since it was announced that chaos improved communication performance in European local fibre network \cite{argyris2005chaos}, more efforts turned to using chaos properties to improve the practical communication system performance. In fact, in the recent decade, various chaos properties fit for communication applications have been reported, including: (1) Chaotic signal is the optimal communication signal allowing for the simplest matched filter (MF) \cite{corron2015chaos}. (2) The Lyapunov spectrum invariance property is used to resist the effect of multipath propagation \cite{ren2013wireless} in wireless communication by eliminating the inter-symbol interference (ISI) caused by multipath \cite{yao2017chaos,yao2019experimental} in theory, and by improving the wireless communication bit error rate (BER) performance using the currently available information and artificial intelligence (AI) \cite{yao2019experimental,Ren2020performance,yin2021direct,yin2021echo,ren2021artificial} in practice. (3) Based on the shape-forming filter (CSF) \cite{renchaotic, yao2019experimental,bai2019experimental}, a chaotic sequence encoding information can be used to replace the conventional chaotic signals generated by the Logistic map and the Lorenz system, providing the extra benefits like higher data transmission rate and lower BER \cite{bai2020double,ren2021cross,bai2021double} in the variant DCSK systems.

A new property of chaos is reported in this Letter, which is referred to as the autocorrelation invariance of the chaotic signal encoding information bits with respect to the base function of the CSF that is used to encode information. We derive the analytical relation about the ACF of the received signal and the wireless channel parameters when the ACF of the transmitted signal is known by the new finding. With this property, the wireless communication channel information can be identified without any probe data in front of the data frame. By this way, the valuable band width in wireless communication system is preserved. Due to the noise immune property of the autocorrelation operation, the channel parameters identification accuracy using the uncovered new property of chaos is improved as compared to the existing channel information blind estimation methods.

To explore the autocorrelation invariance property of chaos, the chaotic baseband wireless communication system (CBWCS) \cite{Ren2020performance} is considered as an example in Fig. 1.

\begin{figure}[ht]
  \centering
  \includegraphics[width=2.8in,height=1.3in]{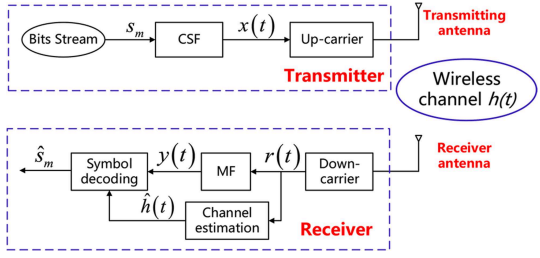}
  \caption{Framework of the CBWCS}
  \label{01}
\end{figure}

In Fig. 1, at the transmitter, the information bit stream ${s_m} \in \left[ { - 1,1} \right]$
is fed into the CSF to generate the chaotic baseband signal $x\left( t \right)$ given by
\begin{equation}
 x\left( t \right) = \sum\limits_{ m= - \infty }^\infty  {{s_m} \cdot p\left( {t - {m \mathord{\left/
 {\vphantom {m f}} \right.
 \kern-\nulldelimiterspace} f}} \right)}
,\end{equation}\\
where $p\left( t \right)$ is the base function of CSF given by

\begin{equation}
 p\left( t \right)\! =\! \left\{ \begin{aligned}&{l}
\left( {1\! -\! {e^{ - {\beta  \mathord{\left/
 {\vphantom {\beta  f}} \right.
 \kern-\nulldelimiterspace} f}}}} \right){e^{\beta t}}\left( {\cos \omega t \!- \!\frac{\beta }{\omega }\sin \omega t} \right), \left( {t < 0} \right)\\
&1\! - \!{e^{\beta \left( {t - {1 \mathord{\left/
 {\vphantom {1 f}} \right.
 \kern-\nulldelimiterspace} f}} \right)}}\left( {\cos \omega t\! -\! \frac{\beta }{\omega }\sin \omega t} \right), \left( {0 \le t < {1 \mathord{\left/
 {\vphantom {1 f}} \right.
 \kern-\nulldelimiterspace} f}} \right)\\
&0, \left( {t \ge {1 \mathord{\left/
 {\vphantom {1 f}} \right.
 \kern-\nulldelimiterspace} f}} \right)
\end{aligned} \right.,
\end{equation}\\
where $\omega  = 2\pi f$ and $0 < \beta  \le f\ln 2$ are the parameters of the base function, $f$ is the base frequency.

\textbf{Discussion 1:} The autocorrelation invariance property of chaos revealed in the specified system configuration in Fig. 1 is a general one that does not depend on the configuration in Fig. 1, which also validates in the cases where the CSF can be used as reported in \cite{cai2021towards,joshi2021synchronization,bai2019digital,bai2018chaos,ren2017chaotic,bai2020double,ren2021cross,bai2021double}.

From Eqs. (1) and (2), the ACF of the transmitting signal is
\begin{equation}
\begin{aligned}
{R_{xx}}\left( \eta  \right) &= \int_{ - \infty }^\infty  {x\left( {t + \eta } \right)\bar x\left( t \right)dt} \\
 &= \int_{ - \infty }^\infty  {\left( {\sum\limits_{m =  - \infty }^\infty  {{s_m}p\left( {t + \eta  - m} \right)} } \right)} \\
 &\times \left( {\sum\limits_{m =  - \infty }^\infty  {{s_m}p\left( {t - m} \right)} } \right)dt\\
 &= \int_{ - \infty }^\infty  {\sum\limits_{i = j = - \infty}^m {{s_i}{s_j}p\left( {t + \eta  - i} \right)p\left( {t - j} \right)dt} } \\
 &+ \int_{ - \infty }^\infty  {\sum\limits_{i = - \infty}^m {\sum\limits_{\scriptstyle j = - \infty\hfill\atop
\scriptstyle j \ne i\hfill}^m {{s_i}{s_j}p\left( {t + \eta  - i} \right)p\left( {t - j} \right)dt} } },
\end{aligned}
\end{equation} where $\eta$ is used to represent the correlation time lag, $\bar x\left( t \right)$  represents the complex conjugate of $x\left( t \right)$, which is replaced by $x\left( t \right)$ because of its real nature.

Considering the independence of the transmitted sym-\\bols, we have $\int_{ - \infty }^\infty  {\sum\limits_{i = - \infty}^m {\sum\limits_{\scriptstyle j = - \infty\hfill\atop
\scriptstyle j \ne i\hfill}^m {{s_i}{s_j}p\left( {t + \eta  - i} \right)p\left( {t - j} \right)dt} } }= 0$. For ${s_i} = {s_j}$, we have ${s_i}{s_j} = 1$. Then equation (3) can be rewritten as
\begin{equation}
{R_{xx}}\left( \eta  \right) = \int_{ - \infty }^\infty  {\sum\limits_{i = j = - \infty}^m {p\left( {t + \eta  - i} \right)p\left( {t - j} \right)dt} }.
\end{equation}

We define $t - i = t - j = \xi $, since $ i = j$, then equation (4) is rewritten as
\begin{equation}
{R_{xx}}\left( \eta  \right) = \int_{ - \infty }^\infty  {p\left( {\xi  + \eta } \right)p\left( \xi  \right)d\xi}.
\end{equation}

From Eq. (5), we conclude that the ACF of the transmitting signal, $x\left( t \right)$, at the transmitter, is precisely the autocorrelation of the base function of CSF as given by
\begin{equation}
 \begin{aligned}
{{R_{xx}}\left( \eta  \right)} = \left\{ \begin{array}{l}
1 + \left( {1 - {e^{ - \beta }}} \right)\frac{{{\omega ^2} - 3{\beta ^2}}}{{2\beta \left( {{\omega ^2} + {\beta ^2}} \right)}}, \eta  = 0\\

{e^{\left( {1 - |\eta |} \right)\beta }}\left( {1 - {e^{ - \beta }}} \right)\frac{{1 - {I_0}}}{2}, \eta  \ne 0
\end{array} \right.
\end{aligned},
 \end{equation}where the parameters $\beta$ and $f$ are defined in Eq. (2), and ${I_0}$ is the autocorrelation at $\eta  = 0$.

This fundamental finding implies that the ACF of the transmitted chaotic signal is available at the receiver, no matter what information symbols are encoded into the chaotic signal. In the following, we investigate the analytical relation between the ACF of the received signal and the wireless channel parameters.

The impulse response $h\left( t \right)$ of the wireless multipath channel is given by
\begin{equation}
  h\left( t \right) = \sum\limits_{l = 0}^{L - 1} {{\alpha _l}\delta \left( {t - {\tau _l}} \right)},
\end{equation}
where ${\alpha _l}$ and ${\tau _l}$ are the attenuation and propagation delay corresponding to path $l$ from the transmitter to the receiver, ${\alpha _l} = {e^{ - \gamma {\tau _l}}}$, $\gamma$ is the damping coefficient \cite{dottling2009radio}, and $\delta \left(  \cdot  \right)$ is the Dirac delta function. It is a statistical average channel model for a practical wireless communication channel. We assume that the delay ${\tau _l}\left( {l = 0,1, \ldots ,L - 1} \right)$ satisfies $0 = {\tau _0} < {\tau _1} <  \cdots  < {\tau _{L - 1}}$.

After the chaotic signal encoded the information symbols by the CSF transmits through the wireless channel, the received signal is given by
\begin{equation}
  r\left( t \right) = h\left( t \right) * x\left( t \right) + w\left( t \right),
\end{equation}
where `$ * $' denotes the convolution, and $w\left( t \right)$ is an additive white Gaussian noise (AWGN).

The ACF of the received signal is given by
\begin{equation}
\begin{aligned}
{R_{rr}}\left( \eta  \right) &= \int_{ - \infty }^\infty  {r\left( {t + \eta } \right)\bar r\left( t \right)dt} \\
 &= \sum\limits_{l = 0}^{L - 1} {\alpha _l^2{R_{xx}}\left( \eta  \right)}  + {R_{ww}}\left( \eta  \right)\\
 &+ \sum\limits_{l = 1}^{L - 1} {{\alpha _0}{\alpha _l}\left[ {{R_{xx}}\left( {\eta  + {\tau _l}} \right) + {R_{xx}}\left( {\eta  - {\tau _l}} \right)} \right]} \\
 &+ \sum\nolimits_{\scriptstyle{l_i},{l_j} = 1\hfill\atop
\scriptstyle{l_i} \ne {l_j}\hfill}^{L - 1} {{\alpha _{{l_i}}}{\alpha _{{l_j}}}{R_{xx}}\left( {\eta  + {\tau _{{l_i}}} - {\tau _{{l_j}}}} \right)},
\end{aligned}
\end{equation}where ${\alpha _l}$, ${\tau _l}$ are defined in Eq. (7).

By examining Eq. (9), we know that the first two terms are the ACFs of the transmitting chaotic signal and the noise, respectively. The third term is the autocorrelation of the transmitting signal with double sides shift corresponding to multipath time delay, while the last term is the autocorrelation of the transmitting signal with double sides shift corresponding to the delay difference of any two multipaths except the main path. Because $w\left( t \right)$ is an AWGN, ${R_{ww}}\left( \eta  \right) = 0$ when $\eta  \ne 0$, and ${R_{ww}}\left( 0  \right)$ corresponds to the variance of noise $w$ \cite{garcia2010robust}.

\begin{figure}[ht]
  \centering
  \includegraphics[width=2.6in,height=1.8in]{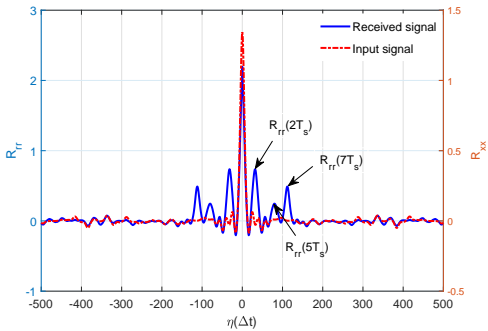}\\
  \caption{The ACF of the received signal $r\left( t \right)$ after multipath propagation.}
  \label{02}
\end{figure}

To demonstrate this conclusion, the ACF of the received signal with three paths is given in Fig. 2, where ${\tau _0} = 0$, ${\tau _1} = 2{T_s}$, ${\tau _2} = 7{T_s}$, $\gamma  = 0.6$. From Fig. 2, it can be seen that, three peaks appear at  ${\tau _1} = 2{T_s}$, ${\tau _2} - {\tau _1} = 5{T_s}$, and ${\tau _2} = 7{T_s}$ in the ACF of the received signal, except for the main path at time delay ${\tau _0} = 0$.

From Eq. (9), we know that the ACF of the received signal, ${R_{rr}}$, is determined by ${R_{xx}}$, and the channel parameters, ${\tau _l}$ and ${\alpha _l}$. The ACF of the received signal can be calculated at the receiver together with the known ACF of the transmitted signal, by this way, the channel parameters can be obtained analytically. Assuming that the delay resolution is equal to one symbol period ${T_s}$, and the maximum delay in the practical wireless communication scenario is $M{T_s}$, then there are at most $L=M+1$ multipaths. The corresponding time delays can be expressed as ${\tau _0} = 0,{\tau _1} = {T_s}, \cdots , {\tau _{L - 1}} = M{T_s}$, and the corresponding channel attenuations are ${\alpha _0},{\alpha _1}, \cdots ,{\alpha _{L - 1}}$. In such a case, substitute the correlation lag $\eta$ in Eq. (9) by $0,{T_s}, \cdots ,k{T_s}, \cdots ,M{T_s}$, respectively. We get the analytical relationship between the channel parameters and the ACFs of the received signal given by Eq. (10) (at the top of next page). In Eq. (10), the ${R_{rr}}\left( {{\tau _0}} \right), \cdots ,{R_{rr}}\left( {k {T_s}} \right), \cdots ,{R_{rr}}\left( {M {T_s}} \right)$ on the left-hand side of the equations represent the ACF values of the received chaotic signal at $\eta= {0}, \cdots ,{k {T_s}}, \cdots ,{M {T_s}}$; on the right-hand side of the equations consist of the channel parameters and the ACF of the transmitting chaotic signal. To this end, we have derived the analytical relationship between the ACF of the received signal and the channel parameters together with the known ACF of the transmitted signal.

\begin{figure*}
\begin{equation}
\begin{aligned}
\left\{ {\begin{array}{*{20}{l}}
{{R_{rr}}\left( 0 \right) = {R_{xx}}\left( 0 \right)\left( {1 + \alpha _1^2 +  \cdots  + \alpha _{L - 1}^2} \right) + {R_{ww}}\left( 0 \right) + \sum\limits_{l = 1}^{L - 1} {{R_{xx}}\left( {{\tau _l}} \right)\left( {2{\alpha _l} + \sum\limits_{{l_i} = 1}^{L - 1 - l} {2{\alpha _{{l_i}}}{\alpha _{{l_i} + l}}} } \right)} }\\
 \vdots \\
{{R_{rr}}\left( {k{T_s}} \right) = {R_{xx}}\left( 0 \right)\left( {{\alpha _k} + \sum\limits_{{l_i} = 1}^{L - 1 - k} {{\alpha _{{l_i}}}{\alpha _{{l_i} + k}}} } \right) + \sum\limits_{l = 1}^{L - 1} {{R_{xx}}\left( {{\tau _l}} \right)\left( {{\alpha _{|k - l|}} + \sum\limits_{{l_i} = 1}^{L - 1 - |k - l|} {{\alpha _{{l_i}}}{\alpha _{{l_i} + |k - l|}}}  + {\alpha _{k + l}} + \sum\limits_{{l_i} = 1}^{L - 1 - k - l} {{\alpha _{{l_i}}}{\alpha _{{l_i} + k + l}}} } \right)} }\\
 \vdots \\
{{R_{rr}}\left( {M{T_s}} \right) = {\alpha _M}{R_{xx}}\left( 0 \right) + \sum\limits_{l = 1}^{L - 1} {{R_{xx}}\left( {{\tau _l}} \right)\left( {{\alpha _{M - l}} + \sum\limits_{{l_i} = 1}^l {{\alpha _{{l_i}}}{\alpha _{{l_i} + M - l}}} } \right)} }
\end{array}} \right..
\end{aligned}
\end{equation}
\end{figure*}

Given Eq. (10), we can solve channel attenuation using ``fsolve'' function in MATLAB \cite{matlab2007young}. The estimated non-zero $\alpha _l$ indicates that the corresponding multipath $l$ exists, and the corresponding time delay is $\tau_l$.


To measure the channel parameters estimation performance, the mean squared error (MSE) criterion is defined as
\begin{equation}
 MS{E_H} = \frac{1}{D}\frac{{\sum\nolimits_{d = 1}^D {||{{\bf{H}}_d} - {{{\bf{\hat H}}}_d}|{|^2}} }}{L},
\end{equation}
where ${\bf{H}} = \left[ {{\alpha _1}, \cdots ,{\alpha _{L - 1}}} \right]$ is the actual channel parameters vector, ${\bf{\hat H}} = \left[ {{{\hat \alpha }_1}, \cdots ,{{\hat \alpha }_{L - 1}}} \right]$ is the estimated channel parameters, $D$ is the number of trials tested, and $L$ is the multipath number.


\begin{figure}[ht]
  \centering
  \includegraphics[width=2.8in,height=2.0in]{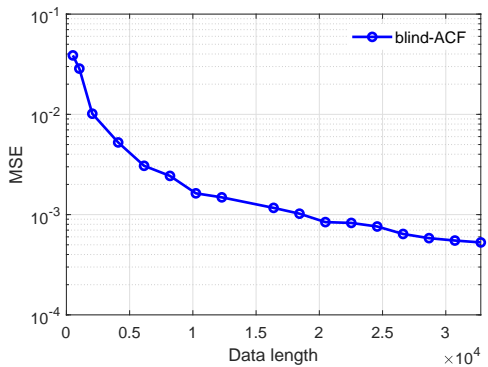}\\
  \caption{Effect of the received data length used for channel parameters estimation in the method based on the finding in this letter.}
  \label{03}
\end{figure}

We investigate the effect of the received signal length used to calculate the ACF on the performance of the channel parameters estimation. The MSE performance versus the data length is plotted in Fig. 3, where the signal to noise ratio (SNR) is 10 dB, and the multipath number $L = 6$. From Fig. 3, it can be seen that the estimation error decreases with the increase of the data length. However, the increasing data length increases computational load accordingly. When the data length $> 2048*16=32768$, the decreasing rate of MSE is small.

Simulations under different channel parameters are performed. The proposed channel parameter estimation method is compared with blind minimum nonlinear prediction error (MNPE) method \cite{zhu2002identification}, blind minimum phase space volume (MPSV) method \cite{leung1998system}, the improved maximum likelihood phase space volume (ML-PSV) method \cite{mukhopadhyay2021blind}, and the optimal non-blind least squared (LS) with white Gaussian driven signal \cite{leung2000blind}. Moreover, the non-blind least squared (LS) method using chaos generated by the CSF is also used for comparison. In the simulations, the channel damping coefficient $\gamma$ is assumed to be a uniformly distributed random variable in the interval [0.3, 0.9], and 100 different channel parameters sets are tested. $1024 * 16$ data size of the received signal for each channel parameter set are used for all comparison methods. The comparison result in Fig. 4 shows the best performance is achieved by the proposed blind estimation method as compared to the blind identification methods including ML-PSV \cite{mukhopadhyay2021blind}, MPSV \cite{leung1998system}, and MNPE \cite{zhu2002identification}. From Fig. 4, we learn that the MSE curve of the proposed method is the closest one to the non-blind LS method with Gaussian driving signal that is optimal one (in theory) in the white noise condition. However, the non-blind LS method needs the probe symbols, which does not belong to the blind estimation methods.

\begin{figure}[ht]
  \centering
  \includegraphics[width=2.8in,height=2.0in]{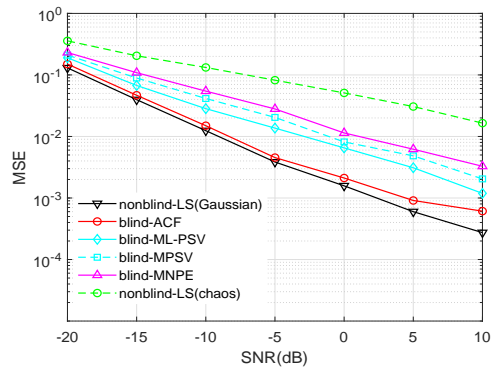}\\
  \caption{The MSE performance comparison result. The black solid line with lower triangular markers is the result using the optimal (non-blind) LS method with Gaussian driving signal, the red solid line with circle markers is the result of the blind estimation method using the new property of chaos here, the cyan solid line with diamond markers is the result using the blind ML-PSV method, the cyan dashed line with squares is the result using the blind MPSV method, the magenta solid line with upper triangular markers is the result using the blind MNPE method, and the green solid line with circle markers is the result using non-blind LS method with chaotic driving signal.}
  \label{04}
\end{figure}

\textbf{Discussion 2:} As shown in Fig. 4, the performance of the non-blind LS using chaos and the blind MNPE, MPSV, and ML-PSV methods are very sensitive to noise, therefore, they get worse MSE performance at low SNR. The blind estimation method using the new property of chaos reported here is not sensitive to noise. Especially, in the low SNR case it achieves very close performance to the optimal one. The reason behind this pleasing phenomenon is that the autocorrelation operation in ACF calculation is robust to the channel noise.

In summary, a fundamental finding that the ACF of chaotic information-encoded signal is identical to the ACF of the base function of CSF, referred to as the autocorrelation invariance property of chaos, is proved analytically in this Letter. This important chaos property allows for blind wireless channel parameters identification. We derive the relationship between the ACF of the received signal and the channel parameters under the known ACF of the transmitted signal. We demonstrate a direct application of the fundamental finding by proposing a novel blind channel information identification method, and show the superior performance of the novel method by both qualitative and quantitative way. As a natural signal, chaos properties promise to improve the performance and efficiency of the conventional wireless communication system. We have explored the application of the new property in non-orthogonal multiple access (NOMA) in multiple input multiple output (MIMO) wireless communication system to demonstrate the improved performance which is not included in this Letter because of length limit.
\bibliography{prl_letter}

\end{document}